\documentclass[a4paper,12pt]{article}
\usepackage[margin=2cm]{geometry}
\usepackage{comment}
\usepackage{amssymb,extarrows,graphicx,subfigure,setspace}
\usepackage{cite}
\usepackage{slashed}
\usepackage{tensor}
\usepackage[toc,page]{appendix}
\usepackage{color}
\usepackage{hyperref}
\usepackage{dirtytalk}
\hypersetup{colorlinks=true, linkcolor=blue, citecolor=red, linktoc=page}
\makeatother

\usepackage{amsmath}
\medskip
\newmuskip\pFqmuskip

\newcommand*\pFq[6][8]{%
  \begingroup 
  \pFqmuskip=#1mu\relax
  \mathcode`\,=\string"8000
  \begingroup\lccode`\~=`\,
  \lowercase{\endgroup\let~}\pFqcomma
  {}_{#2}F_{#3}{\left[\genfrac..{0pt}{}{#4}{#5};#6\right]}%
  \endgroup
}
\newcommand{\pFqcomma}{\mskip\pFqmuskip}

\usepackage{hyperref}

\newcommand{\be}{\begin{equation}}
\newcommand{\bea}{\begin{eqnarray}}
\newcommand{\eea}{\end{eqnarray}}
\newcommand{\ba}{\begin{array}}
\newcommand{\ea}{\end{array}}
\newcommand{\ee}{\end{equation}}
\newcommand{\bes}{\begin{equation*}}
\newcommand{\beas}{\begin{eqnarray*}}
\newcommand{\eeas}{\end{eqnarray*}}
\newcommand{\bas}{\begin{array*}}
\newcommand{\eas}{\end{array*}}
\newcommand{\ees}{\end{equation*}}

\numberwithin{equation}{section}
\begin{document}
	\onehalfspacing
	\noindent
	
	\begin{titlepage}
		\vspace{10mm}
		
		
		\vspace*{20mm}
		\begin{center}
			

{\Large {\bf Modified gravity theories from the Barrow hypothesis}}

			\vspace*{15mm}
			\vspace*{1mm}
		{\bf \large Ankit Anand}\footnote{ E-mail : ankitanandp94@gmail.com \;\;;\;\; anand@physics.iitm.ac.in} 
		\vskip 0.5cm
		{\it
				Department of Physics, Indian Institute of Technology Madras, Chennai 600 036, India}\\
			\vspace{0.2cm}

			\vspace*{1cm}

   {\bf \large Ruben Campos Delgado}\footnote{E-mail :
ruben.camposdelgado@gmail.com} 
		\vskip 0.5cm
		{\it Bethe Center for Theoretical Physics, \\ Physikalisches Institut der Universit\"at Bonn, Nussallee 12, 53115 Bonn, Germany
				}\\
			\vspace{0.2cm}

			\vspace*{1cm}
		\end{center}
		
\begin{abstract}
Barrow proposed that quantum gravity effects might introduce fractal corrections to the area of the event horizon of black holes. The area law gets modified as $S \propto A^{1+\Delta/2}$, with $0\leq\Delta\leq 1$. It was so far unclear whether this assumption could lead to meaningful quantum gravity theories beyond general relativity. In this paper, we argue that this is indeed the case. In particular, assuming $\Delta$ to be a radial function, we show that the Barrow hypothesis, together with the Jacobson's approach can generate non-trivial modified gravity theories.  
\end{abstract}
\end{titlepage}

\newpage
\tableofcontents

\section{Introduction}\label{Sec: Introduction}
Black holes are intriguing entities for several reasons. Hawking's discovery that black holes have a radiation spectrum comparable to that of a black body gives them a perfect laboratory for studying the interactions between quantum mechanics, gravity, and thermodynamics. The concept of Bekenstein-Hawking entropy, often known as black hole entropy, has gained popularity in recent decades. The exploration of black hole entropy is essential for gaining insights into the intricate microscopic aspects of quantum gravity. The renowned Bekenstein–Hawking entropy, which is proportional to the area of the horizon, represents a foundational expression \cite{Maldacena:1997de,Solodukhin:2011gn} in this realm. However, within the broader context of quantum gravity, this entropy is regarded merely as the tree-level outcome of black hole entropy, with subsequent refinements revealing a leading-order quantum gravitational correction in the form of a logarithmic term. This logarithmic correction is deemed fundamental and universal due to its derivation from diverse methods. These approaches encompass conical singularity and entanglement entropy \cite{Solodukhin:2011gn,Solodukhin:1994yz}, Euclidean action method\cite{Fursaev:1994te,Sen:2012dw,El-Menoufi:2015cqw,El-Menoufi:2017kew}, conformal anomaly\cite{Cai:2009ua}, Cardy formula\cite{Carlip:2000nv}, quantum tunneling\cite{Banerjee:2008cf,Banerjee:2008fz}, and quantum geometry\cite{kaul}. This ubiquity underscores the significance of the logarithmic correction in our comprehension of black hole entropy. 

The exploration of black hole entropy is fundamental to grasp the intricate nature of the microscopic constituents within the framework of quantum gravity. One of the most renowned expressions in this realm is the Bekenstein-Hawking entropy, denoted as 
\begin{equation*}
    S_{BH}=\frac{A}{4G},
\end{equation*}
where $A$ is the event horizon area of the black hole. This formula represents a pivotal link between the geometric properties of black holes, such as their event horizon area, and the underlying quantum structure, providing invaluable insights into the nature of spacetime at the smallest scales. The Bekenstein-Hawking entropy differs from conventional thermodynamics in that it is proportional to the black hole horizon area instead of the volume. In classical thermodynamics, a system entropy is proportional to its mass and volume, indicating an extensive and additive quantity. However, the cause for black hole entropy being nonextensive is unknown \cite{Tsallis:2012js}. Based on the Bekenstein-Hawking entropy, there are several developments of alternative construction based on nonextensive statistics, such as Renyi \cite{Renyi} and Tsallis \cite{Tsallis:1987eu}. Recent studies on entropy include Barrow's entropy \cite{Barrow:2020tzx}, Sharma-Mittal's \cite{SayahianJahromi:2018irq}, and Kaniadakis' \cite{Kaniadakis:2005zk, Drepanou:2021jiv}. The first law of thermodynamics links entropy, temperature, internal energy, and heat transport. Changing the definition of entropy can affect these other numbers, as in \cite{Nojiri:2021czz}.

As we don't yet have a complete theory of quantum gravity that works for all energy scales, using effective field theory (EFT) techniques on general relativity gives us a reliable theory of quantum gravity up to an energy scale near the Planck scale \cite{Donoghue:1994dn}. By applying EFT methods, we can recognize certain common aspects of quantum gravity. One of the most fascinating features is the dynamic non-locality of space-time caused by quantum effects when we look at very short distances. Quantum corrections are expected to adjust the black hole metric to match the corrected entropy since it is calculated by the black hole horizon area and event horizon radius. The starting point for the EFT approach is a curvature expansion from which equations of motion can be derived.  Previous studies found that the metric of a Schwarzschild black hole acquires quantum corrections at the third level of curvature \cite{Calmet:2021lny}, while the metric of a Reissner-Nordstr\"om black hole receives modifications already at second order \cite{Delgado:2022pcc}. The authors employed the same strategy to derive a revised geometry and corrected entropy using the Wald formula \cite{Wald:1993nt}, despite various methods available \cite{Cano:2019ycn, Yoon:2007aj, Akbar:2003mv, Sadeghi:2014zna}. 

Barrow \cite{ Barrow:2020tzx} proposed a fractal structure for the horizon of BHs, ironically inspired by the COVID-19 virus form. This structure has a limited volume but an infinite area.  With this proposal, the entropy of the horizon area could potentially exhibit a fractal correction in such a way that the entropy $ S \propto A^{1+\frac{\Delta}{2}}$ as a result of quantum gravitational influences. The parameter $\Delta$, ranging from $0$ to $1$, quantifies the departure from the conventional area law.  In the limit $\Delta \rightarrow 0$, the Barrow entropy becomes the ordinary Bekenstein-Hawking entropy, whereas $\Delta \rightarrow 1$ indicates maximum deformation. Astrophysical observations place an upper bound on $\Delta$ of order $0.03$ \cite{Vagnozzi:2022moj}. The authors of \cite{DiGennaro:2022grw} showed that, despite the fractal correction, the resulting theory of gravity does not differ from general relativity apart from a re-scaled cosmological constant. By applying the Jacobson's approach \cite{Jacobson:1995ab}, consisting of deriving the field equations starting from the form of the entropy, they arrived at
\begin{equation}\label{Modified Einstein with Delta constant}
    R_{\mu \nu}-\frac{1}{2}Rg_{\mu \nu} +\frac{2 \Lambda}{(2+\Delta)A^\frac{\Delta}{2}}  g_{\mu \nu} = 8 \pi G_N T_{\mu \nu}.
\end{equation}
The same authors argued in \cite{DiGennaro:2022ykp} that, in cosmological settings, $\Delta$ should be scale dependent. Remaining in the realm of cosmology, other authors introduced a concept of generalized entropy, of which Barrow is a special case, and computed the corrections to the Friedmann equation \cite{Nojiri:2022aof, Nojiri:2022dkr}. Inspired by these previous works, in this paper we consider the case where $\Delta$ is not constant, but rather it is a function of the radial coordinate, i.e.  $\Delta=\Delta(r)$. We show that this assumption is powerful enough to generate modified gravity theories beyond general relativity. 

The paper is organized as follows. In Section \ref{sec: motivation}, we argue once and for all that $\Delta$ can not be a constant and we motivate our choice of considering it a function of the radial distance $r$. In Section \ref{Sec:Corrected Einstein Equation}, we follow the Jacobson's approach to derive the modified Einstein field equations due to small deviations from constant $\Delta$. We present the equations up to second order in the expansion parameter. In Section \ref{sec:metric correction}, we solve the equations for a fixed plane $\theta=\frac{\pi}{2}$ and show that they generalize several results already present in the literature. Finally, in section \ref{Sec:Conclusion}, we summarise our results and discuss possible future directions. 
\section{Why do we need a non-constant \texorpdfstring{$\Delta$}{TEXT}?}\label{sec: motivation}
Before beginning with our calculation, it is a good idea to step back for a while and comment about the nature of $\Delta$ and its radial functional dependence. Firstly, we would like to clarify a misconception which is present in the current literature. The authors of \cite{Jusufi:2021fek,Vagnozzi:2022moj}, starting from the expression of the entropy, computed the Hawking temperature as $\frac{1}{T}=\frac{\partial S}{\partial M}$ and obtained (with $G_N=1$)
\begin{equation}\label{eq:wrong_temp}
    T=\frac{1}{(\Delta+2)(4\pi)^{1+\Delta/2}M^{1+\Delta}}.
\end{equation}
From that, they used the formula $T=\frac{f'(r)}{4\pi}|_{r=r_h}$ and got the $g_{tt}=g_{rr}=f(r)$ component of the metric:
\begin{equation}\label{eq:wrong_f}
    f(r)=1-\frac{(\Delta+2)M^{\Delta+1}(4\pi)^{\frac{\Delta}{2}}}{r}.
\end{equation}
Naively, it seems that a constant $\Delta$ leads to a modified Schwarzschild solution. However, this is not the case. In fact, as already pointed out in \cite{DiGennaro:2022grw}, Eq. \eqref{eq:wrong_f} is just a re-definition of the Schwarzschild metric with ADM mass $\mathcal{M}$ equal to
\begin{equation}
    \mathcal{M}=\frac{1}{2}(\Delta+2)(4\pi)^{\Delta/2}M^{\Delta+1}. 
\end{equation}
Indeed, when one observes a black hole and measures its mass, one actually measures $\mathcal{M}$ and not $M$ which remains a not observable quantity. Moreover, plugging $\mathcal{M}$ into Eq. \eqref{eq:wrong_temp} yields the usual expression of the temperature $T=1/(8\pi M)$. There is also a more formal reason of why Eq. \eqref{eq:wrong_f} is not a modified Schwarzschild solution. All theories of quantum gravity which provide an explicit quantum correction to the metric predict that $g_{rr}\neq g_{tt}$, see e.g. \cite{Xiao:2021zly, Calmet:2021lny, Delgado:2022pcc}. It is then natural to assume that this remains true also for the Barrow hypothesis. 

Hence, in order to generate modified gravity theories beyond general relativity, $\Delta$ can not be a constant and must depend on the parameters defining the problem at hand. For a Schwarzschild black hole, the physical quantities of interest are the radial distance $r$ and the angle $\theta$. If we restrict ourselves to a fixed plane, say $\theta=\pi/2$, then  $\Delta$ has to depend on $r$ only, i.e. $\Delta=\Delta(r)$.  Another reason for the radial dependence comes from analyzing the series expansion of $S\propto A^{1+\Delta/2}$. Up to first order in $\Delta$ we have
\begin{equation}
    S\propto \frac{A}{4}+\frac{A}{8}\ln\frac{A}{4}\Delta.
\end{equation}
In order to keep the subleading term small at $\mathcal{O}(1)$, $\Delta$ should scale inversely proportional to $A\ln A$. This guarantees that the quantum gravity correction is small when the black hole is large (i.e. when the energy scale is small). A third reason in support of the radial dependence is motivated by some important results recently obtained in cosmology. The authors of \cite{DiGennaro:2022ykp} proposed an exponential decay for $\Delta$ and successfully computed the modified Friedmann equation. They predicted a variation of the sign of the cosmological constant which could solve the still open problem of the Hubble tension. 

In the next section we are going to compute modifications to the Einstein equations by explicitly expanding $\Delta$ around the constant solution. 
\section{Modified Einstein equations with the Jacobson's approach}\label{Sec:Corrected Einstein Equation}
The Barrow hypothesis suggests that the entropy has the form
\begin{equation*}
    S_{\rm BH} = \frac{\mathcal{A}}{4G_N},
\end{equation*}
where $\mathcal{A}=A^{1+\frac{\Delta}{2}}$, and $\Delta$ ranges from $0$ to $1$. For a spherically symmetric black hole of radius $r_h$, $A=4\pi r^2_h$. Using the Raychaudhuri equation:
\begin{equation}
\int \lambda T_{\mu \nu} k^\mu k^\nu d\lambda d \mathcal{A} =  \frac{1}{8 \pi G_N} \int \lambda R_{\mu \nu} k^\mu k^\nu d\lambda d \mathcal{A},
\end{equation}
we arrive at
\begin{equation}
\int \left[T_{\mu \nu}-\frac{1}{8 \pi G_N} R_{\mu \nu}\right] \left(1+\frac{\Delta}{2}\right)  k^\mu k^\nu A^\frac{\Delta}{2} d\lambda d A = 0,
\end{equation}
which means 
\begin{equation}\label{Einstein Equation}
    \left[T_{\mu \nu}-\frac{1}{8 \pi G_N} R_{\mu \nu}\right] \left(1+\frac{\Delta}{2}\right) A^\frac{\Delta}{2} = \mathfrak{f}\;g_{\mu \nu}
\end{equation}
for some function $\mathfrak{f}$.
Taking the covariant derivative of Eq. $\eqref{Einstein Equation}$ we eventually obtain
\begin{equation}\label{F function}
    -\frac{1}{16 \pi G_N} \nabla_\nu R \left(1+\frac{\Delta}{2}\right)A^{\frac{\Delta}{2}} +\mathfrak{f} \frac{ \nabla_\nu \Delta }{2+\Delta} \left[1+\left(1+\frac{\Delta}{2}\right)\ln{A}\right] = \nabla_\nu \mathfrak{f}.
\end{equation}
To get \eqref{F function}, we used the assumption of stationary horizon $\nabla^{\mu}A=0$, as well as the Bianchi identity $\nabla^{\mu}R_{\mu\nu}=\frac{1}{2}\nabla_{\nu}R$. 
Assuming $\mathfrak{f}$ to be a function of $r$ only, we see that it satisfies the differential equation
\begin{equation}\label{F function equation}
      \mathfrak{f} = \frac{\Delta+2}{\partial_r \Delta \left[1+\left(1+\frac{\Delta}{2}\right) \ln{A}\right]} \left\{\partial_r \mathfrak{f}+\frac{(\Delta+2)A^{\frac{\Delta}{2}}}{32 \pi G_N}\partial_r R\right\}.  
\end{equation}
The general solution of Eq. \eqref{F function equation} is 
\begin{equation}\label{F(r)}
    \mathfrak{f}(r) = -\frac{(\Delta (r)+2)^{1+\left(\frac{\Delta (r)}{2}+1\right)\ln{A} }}{32 \pi  G_N}  \int^r du\, A^{\frac{\Delta (r)}{2}}  (\Delta (r)+2)^{-\left(\frac{\Delta (r)}{2}+1\right)\ln{A}} \partial_u R.
\end{equation}
If $\Delta$ is constant, then
\begin{equation}\label{f(r) constant delta}
    \mathfrak{f}(r) =  -\frac{\left(1+\frac{\Delta}{2} \right) A^{\frac{\Delta}{2}}}{8 \pi G_N} \frac{R}{2} + \Lambda',
\end{equation}
where $\Lambda'$ is the integration constant. This expression matches the result of \cite{DiGennaro:2022grw}.
Since our goal is to see what happens when $\Delta$ is a radial function, thus we expand
\begin{equation}\label{Delta(r)}
    \Delta(r) = c + \epsilon\; \kappa(r) \ ,
\end{equation}
where $c$ is some constant. By doing a series expansion of Eq. \eqref{F(r)}, integrating by parts and collecting powers of $\epsilon$, we find
\begin{equation}\label{corrected_Einstein}
\begin{gathered}
     R_{\mu \nu}-\frac{1}{2}Rg_{\mu \nu} + \frac{\epsilon}{4} \ln{A}\int_r^\infty du\, R \partial_u \kappa(u)  g_{\mu \nu} \\
     - \frac{\epsilon^2}{16}  \frac{\ln{A}}{c+2} g_{\mu \nu}\Big\{\left[-4+2(c+2)\ln{A}\ln^2(c+2)\right]
    \int_r^\infty du\, R \kappa(u) \partial_u\kappa(u)\\
    - 2(c+2)\log{A} \ln^2(c+2)\kappa(r)\int_r^\infty du\, R \partial_u\kappa(u)\Big\} + \mathcal{O}(\epsilon^3)= 8 \pi G_N T_{\mu \nu}.
\end{gathered}
\end{equation}
The limit $\epsilon\rightarrow 0$ reduces to general relativity.
\section{Quantum gravitational corrections to the Schwarzschild metric}\label{sec:metric correction}
In order to solve Eq. \eqref{corrected_Einstein}, we start with the Ansatz 
\begin{equation}\label{ansatz}
    ds^2= -f(r) dt^2 + \frac{dr^2}{g(r)} + h(r)\; r^2 \;d\theta^2 + k(r)\; r^2\;\sin^2{\theta} \;d\theta^2 \ ,
\end{equation}
where 
\begin{eqnarray}
    f(r) &=& 1- \frac{2G_N M}{r} + \epsilon \zeta_1(r) + \epsilon^2 \zeta_2(r) + \cdots \ , \nonumber \\
    g(r) &=& 1- \frac{2G_N M}{r} + \epsilon \Sigma_1(r) + \epsilon^2 \Sigma_2(r) + \cdots \ , \nonumber \\
    h(r) &=& 1 + \epsilon \frac{\Omega_1(r)}{r^2} + \epsilon^2     \frac{\Omega_2(r)}{r^2} + \cdots \ , \nonumber \\
     k(r) &=& 1 + \epsilon \frac{\Xi_1(r)}{r^2 \sin^2{\theta} } + \epsilon^2     \frac{\Xi_2(r)}{r^2 \sin^2{\theta} } + \cdots \ . \nonumber
\end{eqnarray}
The Ricci scalar has the structure 
\begin{equation}
    R = \epsilon R_1 + \epsilon^2 R_2 + \cdots \ .
\end{equation}
Hence, at order $\epsilon^2$ the Einstein equations reduce to 
\begin{equation}\label{eq:Einstein_new}
    R_{\mu \nu}-\frac{1}{2}Rg_{\mu \nu} + \frac{\epsilon^2}{4} g_{\mu \nu} \ln{A} \int_r^\infty du\,R_1 \partial_u \kappa(u) = 0.
\end{equation}
It is easy to verify that $\zeta_1=\Sigma_1=\Omega_1=\Xi_1=0$. However, $R_1$ is a function of $\zeta_1,
\Sigma_1, \Omega_1, \Xi_1$, so it is also identically zero. To have non-trivial corrections, we must require 
\begin{equation}
    R_1 \partial_u\kappa(u) = \mathfrak{A}(u),
\end{equation}
for some function $\mathfrak{A}(u) \neq 0$, i.e.
\begin{equation}
  \kappa(u) = \int^u dt\, \frac{1}{R_1} \mathfrak{A}(t).
\end{equation}
Since $R_1$ is zero, we should regularize the above expression as
\begin{equation}
    \kappa(r)  =  \lim_{\delta \to 0} \int^r dt \frac{1}{\delta}\; \mathfrak{A}(t).
\end{equation}
The relation $\lim \limits_{\substack{%
    \epsilon \to 0}}\Delta(r)=c$ is recovered:
\begin{equation}
    \lim \limits_{\substack{%
    \epsilon \to 0}} \Delta(r)= c +\lim \limits_{\substack{%
    \epsilon \to 0\\
   \delta \to 0}} \frac{\epsilon}{\delta} \int^r dt \, \mathfrak{A}(t)=c,
\end{equation}
provided that $\epsilon$ goes to zero faster than $\delta$.

The solutions of the modified Einstein equation \eqref{eq:Einstein_new} depend on the function
   \begin{eqnarray}\label{I(r)}
       \mathcal{I} (r) := \int_r^\infty du\, R_1\;\partial_u \kappa(u) = \int_r^\infty du\; \mathfrak{A}(u).
   \end{eqnarray}
Unfortunately, we do not know exactly what $\mathfrak{A}$ is (and hence $\mathcal{I}(r)$). It cannot be determined by first principles. However, we could try to guess the expression of $\mathcal{I}(r)$ such that we recover the quantum gravitational corrected metrics already present in the literature. For simplicity we set $\theta=\pi/2$ and hence $h(r)=k(r)$. We have three independent equations, corresponding to the $tt$, $rr$, $\theta\theta$ (or $\phi\phi$) components of Eq. \eqref{eq:Einstein_new} which allow us to uniquely determine $\zeta_2$, $\Sigma_2$ and $\Omega_2$. In order to solve the equations, we make use of the xTras package of Mathematica \cite{Nutma:2013zea}. We make an Ansatz for $\mathcal{I}(r)$ with a variable number of coefficients, then determine the expression of these coefficients by comparing the form of our metric with the ones derived in previous works. In particular, we choose to compare our results with the ones of Xiao and Tian \cite{Xiao:2021zly} and Calmet and Kuipers \cite{Calmet:2021lny}, which were obtained via the EFT framework for quantum gravity. 
\subsubsection*{Case I}
Our first Ansatz for $\mathcal{I}(r)$ is
\begin{equation}
    \mathcal{I}(r) = \frac{a_1}{r^4} + \frac{a_2}{r^5} +\frac{a_3}{r^4}\ln{(\mu r)} +\frac{a_4}{r^5}\ln{(\mu r)},
\end{equation}
where $\mu$ is an arbitrary energy scale.
The solutions of Eq. \eqref{eq:Einstein_new} are
\begin{equation}
    \Omega_2(r)=-\frac{\ln{A} \left[6 \ln {(\mu r)}(6r a_3 +5 a_4)+9 r (4 a_1+a_3)+30 a_2+4 a_4\right]}{72 r},
\end{equation}
\begin{equation}
\begin{split}
    \Sigma_2(r) &= \ln{A}\bigg\{\frac{6 \ln (\mu r) \left[G_NM(6a_3 r+15a_4) -r(9a_3r+11a_4)\right]}{72 r^4} \\
    &-\frac{9 r \left[G_NM(-4 a_1 +7a_3) + 6 a_1 r\right]-a_2 (90 G_N M-66 r)-a_4 (17 r-48 G_N M)}{72 r^4}\bigg\},
\end{split}
\end{equation}
\begin{equation}
\begin{split}
   \zeta_2(r) &= -\ln {A} \bigg\{\frac{ 6 \ln (\mu r) \left[3 a_3 r (2 G_N M+r)+a_4 (5 G_N M+r)\right]}{72 r^4} \\
   &+\frac{9 r \left[G_NM(4 a_1 +a_3)+(2 a_1+3a_3)r\right]+6 a_2 (5 G_N M+r)+a_4 (4 G_N M+5 r)}{72 r^4}\bigg\}.
\end{split}
\end{equation}
We compare our solutions with the ones of Xiao and Tian \cite{Xiao:2021zly}, which read
\begin{equation}
\begin{gathered}
    f(r)=1-\frac{2G_NM}{r}+\gamma G^2_N\left(-\frac{512\pi M}{3r^3}+\frac{256\pi M^2}{r^4}\right),\\
    g(r)=1-\frac{2G_N M}{r}+\gamma G^2_N\left(-\frac{256\pi M}{r^3}+\frac{1280\pi M^2}{3r^4}\right).
\end{gathered}
\end{equation}
By comparing the different powers of $r$ we can determine the expressions of the (dimensionful) constants $a_1, a_2, a_3$ and $a_3$:
\begin{equation}\label{coeff_Xiao}
\begin{gathered}
    a_1 = \frac{11456 \pi  \gamma  G_N}{25 \epsilon ^2 \ln {A}}, \hspace{5mm} a_2 = -\frac{31744 \pi  \gamma  G^2_N M}{75 \epsilon ^2 \ln {A}}, \\
     a_3  = \frac{45824 \pi  \gamma  G_N}{75 \epsilon ^2 \ln{A}},\hspace{5mm}a_4= -\frac{7168 \pi  \gamma  G^2_N M}{5 \epsilon ^2 \ln {A}}, 
\end{gathered}
\end{equation}
With all these terms the final expression for the metric is 
\begin{equation}\label{sol1_Xiao}
\begin{split}
    f(r) &= 1-\frac{2G_NM}{r} -\frac{8592 \pi  \gamma  G_N}{25 r^2} -\frac{11456 \pi  \gamma  G_N \ln(\mu r)}{75 r^2}  -\frac{512 \pi  \gamma  G^2_N M}{3 r^3}  \\
    & -\frac{13952 \pi  \gamma  G^2_N M \ln (\mu r)}{75 r^3} +\frac{256 \pi  \gamma  G^3_N M^2}{r^4} + \frac{1792 \pi  \gamma  G^3_N M^2 \ln (\mu r)}{3 r^4},
\end{split}
\end{equation}
\begin{equation}\label{sol2_Xiao}
\begin{split}
    g(r) &=  1-\frac{2G_NM}{r} -\frac{8592 \pi  \gamma  G_N}{25 r^2} - \frac{11456 \pi  \gamma  G_N \ln (\mu r)}{25 r^2}-\frac{256 \pi  \gamma  G^2_N M}{r^3} \\
    &+\frac{121472 \pi  \gamma  G^2_N M \ln (\mu r)}{75 r^3} +\frac{1280 \pi  \gamma  G^3_N M^2}{3 r^4} - \frac{1792 \pi  \gamma  G^3_N M^2 \ln (\mu r)}{r^4},  
\end{split}
\end{equation}
\begin{equation}\label{sol3_Xiao}
 h(r) = 1 -\frac{22912 \pi  \gamma  G_N}{75 r^2} - \frac{22912 \pi  \gamma  G_N \ln (\mu r)}{75r^2} +\frac{256 \pi  \gamma  G^2_N M}{r^3}  + \frac{1792 \pi  \gamma  G^2_N M \ln (\mu r)}{3 r^3}.
\end{equation}
which generalizes the result of \cite{Xiao:2021zly}. Notice that the coefficients \eqref{coeff_Xiao} are formally divergent, but the solutions \eqref{sol1_Xiao}-\eqref{sol3_Xiao} are finite. 


\subsubsection*{Case II}
Another possible choice for $\mathcal{I}(r)$ is
\begin{equation}
    \mathcal{I}(r) = \frac{a_1}{r^7} + \frac{a_2}{r^8} +\frac{a_3}{r^7}\ln{(\mu r)} +\frac{a_4}{r^8}\ln{(\mu r)}.
\end{equation}
The solutions of Eq. \eqref{eq:Einstein_new} are
\begin{equation}
    \Omega_2(r)=-\frac{\ln{A} \left[30 \ln (\mu r)(21 a_3 r+20 a_4)+(630 a_1+36a_3)r +600 a_2+25 a_4\right]}{1800 r^4},
\end{equation}
\begin{equation}
\begin{split}
    \Sigma_2(r) &= \ln{A}\bigg\{\frac{ 9 r \left[-448 a_3 G_N M + 20 a_1 (98 G_N M - 57 r) + 207 a_3 r\right]-4 a_4 (975 G_N M - 464 r)}{7200 r^7} \\
    &+\frac{120 a_2 (180 G_N M - 101 r)+60 \ln (\mu r) \left[3 a_3 r (98 G_N M-57 r)+a_4 (360 G_N M-202 r)\right]}{7200 r^7}\bigg\},
\end{split}
\end{equation}
\begin{equation}
\begin{split}
   \zeta(r) &=-\ln{A} \bigg\{\frac{9 r \left[280 a_1 G_N M + 16 a_3 G_N M + r(20 a_1  + 9 a_3)\right]+4 a_4 (25 G_N M + 11 r)}{7200 r^7} \\
   &+\frac{120 a_2 (20 G_N M + r)+60 \ln(\mu r) \left[3 a_3 r (14 G_N M+r)+2 a_4 (20 G_N M+r)\right]}{7200 r^7}\bigg\}.
\end{split}
\end{equation}
We compare now our solutions with the ones of Calmet and Kuipers, \cite{Calmet:2021lny}, which read
\begin{equation}
\begin{gathered}
    f(r)=1-\frac{2G_N M}{r}+640\pi c_6\frac{G^5_N M^3}{r^7},\\
    g(r)=1-\frac{2G_NM}{r}+128\pi c_6 \frac{G^4_NM^2}{r^6}\left(27-49\frac{G_NM}{r}\right).
\end{gathered}
\end{equation}
By comparing the different powers of $r$ we can find the values of $a_1$,$a_2$,$a_3$ and $a_4$:
\begin{equation}\label{coeff_Calmet}
\begin{gathered}
     a_1 =\frac{75168 \pi  c_6 G^3_N M}{1225 \epsilon ^2 \ln {A}}, \hspace{5mm} a_2 = -\frac{1952 \pi  c_6  G^4_N M^2}{ \epsilon ^2 \ln {A}}\\
  a_3 = \frac{11136 \pi  c_6  G^3_N M}{35 \epsilon ^2 \ln {A}}, \hspace{5mm}a_4= -\frac{768 \pi  c_6 G^4_N M^2}{ \epsilon ^2 \ln {A}}.
\end{gathered}
\end{equation}
With all these terms the final expression for the metric is 
\begin{equation}\label{sol1_Calmet}
\begin{split}
f(r) &=1-\frac{2G_NM}{r} -\frac{6264 \pi c_6 G^3_N  M}{1225 r^5} -\frac{1392 \pi c_6  G^3_N M \ln (\mu r)}{175 r^5} - \frac{3104 \pi c_6  G^4_N  M^2 \ln (\mu r)}{25 r^6} \\
    & +\frac{640 \pi c_6 G^5_N  M^3}{r^7} - \frac{256 \pi c_6 G^5_N  M^3 \ln (\mu r)}{r^7}, 
\end{split}
\end{equation}
\begin{equation}\label{sol2_Calmet}
\begin{split}
g(r) &= 1-\frac{2G_NM}{r}-\frac{6264 \pi c_6 G^3_N  M}{1225 r^5} -\frac{79344 \pi c_6 G^3_N  M \ln (\mu r)}{175 r^5} +\frac{3456 \pi c_6 G^4_N  M^2}{r^6}  \\
    &+  \frac{2304 \pi c_6 G^5_N  M^3 \ln (\mu r)}{r^7}-\frac{6272 \pi c_6 G^5_N M^3}{r^7} -\frac{12832 \pi c_6  G^4_N  M^2 \ln (\mu r)}{25 r^6}.
\end{split}
\end{equation}
\begin{equation}\label{sol3_Calmet}
\begin{split}   
h(r) &= 1 -\frac{696 \pi c_6 G^3_N  M}{25 r^5} -\frac{2784 \pi c_6  G^3_N  M \ln (\mu r)}{25 r^5} +\frac{640 \pi c_6  G^4_N M^2}{r^6}  -\frac{256 \pi c_6 G^4_N M^2 \ln (\mu r)}{r^6},
\end{split}
\end{equation}
which generalizes the result of \cite{Calmet:2021lny}. Again, notice that the coefficients \eqref{coeff_Calmet} are formally divergent, but the solutions \eqref{sol1_Calmet}-\eqref{sol3_Calmet} are finite. 
\section{Conclusions}\label{Sec:Conclusion}
The Barrow hypothesis posits a fractal structure at the black hole horizon which modifies the form of the Bekenstein-Hawking entropy area law according to a parameter $\Delta$. If $\Delta$ is constant, then the Barrow hypothesis does not give any substantial modifications to general relativity. In this paper we showed that the assumption of $\Delta$ as function of the radial distance leads to modified gravity theories beyond general relativity. All these theories have a common feature: they modify the classical Schwarzschild black hole metric with quantum gravitational corrections. Interestingly, we found corrections not only to the $g_{tt}$ and $g_{rr}$ components, but also to the $g_{\theta\theta}$ and $g_{\phi\phi}$ components. This should not be a surprise, since the central point of the Barrow hypothesis is to replace spherical symmetry with fractal symmetry. The different theories that one can generate are encoded in the choice for $\mathcal{I}(r)$ (see \eqref{I(r)}).  However, the specific expression of $\mathcal{I}(r)$ remains indeterminate a priori, i.e. there is no best choice for $\mathcal{I}(r)$ and one can in principle obtain any form of quantum gravitational corrections. Furthermore, we assumed so far that the Barrow hypothesis is connected to already existing theories, but this does not necessarily have to be true. It may be that the Barrow hypothesis leads to completely new, still unknown quantum corrections. In this regard, an interesting future direction of research would be to compute some black hole thermodynamic quantities like temperature and pressure, and see whether the answer matches with the already available results. A similar calculation with constant $\Delta$ was carried out in \cite{Abreu:2024tdv} and within the generalized entropy framework in \cite{Nojiri:2022sfd}. Despite the aforementioned difficulties, we can rightly affirm that the Barrow hypothesis has become a rich and valid framework within the realm of quantum gravity. 





\begin{thebibliography}{99}



\bibitem{Maldacena:1997de}
J.~M.~Maldacena, A.~Strominger and E.~Witten,
\emph{Black hole entropy in M theory,}
JHEP \textbf{12} (1997) 002,
[arXiv:hep-th/9711053 [hep-th]].

\bibitem{Solodukhin:2011gn}
S.~N.~Solodukhin,
\emph{Entanglement entropy of black holes,}
Living Rev. Rel. \textbf{14} (2011) 8,
[arXiv:1104.3712 [hep-th]].

\bibitem{Solodukhin:1994yz}
S.~N.~Solodukhin,
\emph{The Conical singularity and quantum corrections to entropy of black hole,}
Phys. Rev. D \textbf{51} (1995) 609-617,
[arXiv:hep-th/9407001 [hep-th]].

\bibitem{Fursaev:1994te}
D.~V.~Fursaev,
\emph{Temperature and entropy of a quantum black hole and conformal anomaly,}
Phys. Rev. D \textbf{51} (1995) 5352-5355,
[arXiv:hep-th/9412161 [hep-th]].

\bibitem{Sen:2012dw}
A.~Sen,
\emph{Logarithmic Corrections to Schwarzschild and Other Non-extremal Black Hole Entropy in Different Dimensions,}
JHEP \textbf{04} (2013) 156,
[arXiv:1205.0971 [hep-th]].



\bibitem{El-Menoufi:2015cqw}
B.~K.~El-Menoufi,
\emph{Quantum gravity of Kerr-Schild spacetimes and the logarithmic correction to Schwarzschild black hole entropy,}
JHEP \textbf{05} (2016) 035,
[arXiv:1511.08816 [hep-th]].


\bibitem{El-Menoufi:2017kew}
B.~K.~El-Menoufi,
\emph{Quantum gravity effects on the thermodynamic stability of 4D Schwarzschild black hole,}
JHEP \textbf{08} (2017) 068,
[arXiv:1703.10178 [gr-qc]].

\bibitem{Cai:2009ua}
R.~G.~Cai, L.~M.~Cao and N.~Ohta,
\emph{Black Holes in Gravity with Conformal Anomaly and Logarithmic Term in Black Hole Entropy,}
JHEP \textbf{04} (2010) 082,
[arXiv:0911.4379 [hep-th]].

\bibitem{Carlip:2000nv}
S.~Carlip,
\emph{Logarithmic corrections to black hole entropy from the Cardy formula,}
Class. Quant. Grav. \textbf{17} (2000)  4175-4186,
[arXiv:gr-qc/0005017 [gr-qc]].


\bibitem{Banerjee:2008cf}
R.~Banerjee and B.~R.~Majhi,
\emph{Quantum Tunneling Beyond Semiclassical Approximation,}
JHEP \textbf{06} (2008) 095,
[arXiv:0805.2220 [hep-th]].

\bibitem{Banerjee:2008fz}
R.~Banerjee and B.~R.~Majhi,
\emph{Quantum Tunneling, Trace Anomaly and Effective Metric,}
Phys. Lett. B \textbf{674} (2009) 218-222,
[arXiv:0808.3688 [hep-th]].

\bibitem{kaul}
R.~K.~Kaul and P.~Majumdar,
\emph{Logarithmic correction to the Bekenstein-Hawking entropy,}
Phys. Rev. Lett. \textbf{84} (2000) 5255-5257,
[arXiv:gr-qc/0002040 [gr-qc]].

\bibitem{Tsallis:2012js}
C.~Tsallis and L.~J.~L.~Cirto,
\emph{Black hole thermodynamical entropy,}
Eur. Phys. J. C \textbf{73} (2013) 2487,
[arXiv:1202.2154 [cond-mat.stat-mech]].

\bibitem{Renyi}
A. ~Rényi,
\emph{On measures of information and entropy,}
Berkeley Symp. on Math. Statist. and Prob. (1961) 547-561.

\bibitem{Tsallis:1987eu}
C.~Tsallis,
\emph{Possible Generalization of Boltzmann-Gibbs Statistics,}
J. Statist. Phys. \textbf{52} (1988) 479-487,

\bibitem{Barrow:2020tzx}
J.~D.~Barrow,
\emph{The Area of a Rough Black Hole,}
Phys. Lett. B \textbf{808} (2020) 135643,
[arXiv:2004.09444 [gr-qc]].

\bibitem{SayahianJahromi:2018irq}
A.~Sayahian Jahromi, S.~A.~Moosavi, H.~Moradpour, J.~P.~Morais Gra\c{c}a, I.~P.~Lobo, I.~G.~Salako and A.~Jawad,
\emph{Generalized entropy formalism and a new holographic dark energy model,}
Phys. Lett. B \textbf{780} (2018) 21-24,
[arXiv:1802.07722 [gr-qc]].

\bibitem{Kaniadakis:2005zk}
G.~Kaniadakis,
\emph{Statistical mechanics in the context of special relativity. II.,}
Phys. Rev. E \textbf{72} (2005) 036108,
[arXiv:cond-mat/0507311 [cond-mat]].

\bibitem{Drepanou:2021jiv}
N.~Drepanou, A.~Lymperis, E.~N.~Saridakis and K.~Yesmakhanova,
\emph{Kaniadakis holographic dark energy and cosmology,}
Eur. Phys. J. C \textbf{82} no.5, (2022) 449,
[arXiv:2109.09181 [gr-qc]].

\bibitem{Nojiri:2021czz}
S.~Nojiri, S.~D.~Odintsov and V.~Faraoni,
\emph{Area-law versus R\'enyi and Tsallis black hole entropies,}
Phys. Rev. D \textbf{104} no.8, (2021) 084030,
[arXiv:2109.05315 [gr-qc]].

\bibitem{Donoghue:1994dn}
J.~F.~Donoghue,
\emph{General relativity as an effective field theory: The leading quantum corrections,}
Phys. Rev. D \textbf{50} (1994) 3874-3888,
[arXiv:gr-qc/9405057 [gr-qc]].

\bibitem{Calmet:2021lny}
X.~Calmet and F.~Kuipers,
\emph{Quantum gravitational corrections to the entropy of a Schwarzschild black hole,}
Phys. Rev. D \textbf{104} no.6, (2021) 66012,
[arXiv:2108.06824 [hep-th]].

\bibitem{Delgado:2022pcc}
R.~C.~Delgado,
\emph{Quantum gravitational corrections to the entropy of a Reissner\textendash{}Nordstr\"om black hole,}
Eur. Phys. J. C \textbf{82} no.3, (2022) 272,
[erratum: Eur. Phys. J. C \textbf{83} (2023) no.6, 468]
[arXiv:2201.08293 [hep-th]].

\bibitem{Wald:1993nt}
R.~M.~Wald,
\emph{Black hole entropy is the Noether charge,}
Phys. Rev. D \textbf{48} no.8, (1993) R3427-R3431,
[arXiv:gr-qc/9307038 [gr-qc]].

\bibitem{Cano:2019ycn}
P.~A.~Cano, S.~Chimento, R.~Linares, T.~Ort\'\i{}n and P.~F.~Ram\'\i{}rez,
\emph{$\alpha'$ corrections of Reissner-Nordstr\"om black holes,}
JHEP \textbf{02} (2020) 031,
[arXiv:1910.14324 [hep-th]].

\bibitem{Yoon:2007aj}
M.~Yoon, J.~Ha and W.~Kim,
\emph{Entropy of Reissner-Nordstrom Black Holes with Minimal Length Revisited,}
Phys. Rev. D \textbf{76} (2007) 047501,
[arXiv:0706.0364 [gr-qc]].

\bibitem{Akbar:2003mv}
M.~M.~Akbar and S.~Das,
\emph{Entropy corrections for Schwarzschild and Reissner-Nordstr\"om black holes,}
Class. Quant. Grav. \textbf{21} (2004) 1383-1392,
[arXiv:hep-th/0304076 [hep-th]].

\bibitem{Sadeghi:2014zna}
J.~Sadeghi, B.~Pourhassan and F.~Rahimi,
\emph{Logarithmic corrections of charged hairy black holes in (2 + 1) dimensions,}
Can. J. Phys. \textbf{92} no.12, (2014) 1638-1642
[arXiv:1708.07383 [gr-qc]].

\bibitem{Jacobson:1995ab}
T.~Jacobson,
\emph{Thermodynamics of space-time: The Einstein equation of state,}
Phys. Rev. Lett. \textbf{75} (1995) 1260-1263,
[arXiv:gr-qc/9504004 [gr-qc]].

\bibitem{Jusufi:2021fek}
K.~ Jusufi et al.,
\emph{Constraints on Barrow Entropy from M87* and S2 Star Observations,}
Universe \textbf{8} no.2, (2022) 102,
[arXiv:2110.07258 [gr-qc]].

\bibitem{Vagnozzi:2022moj}
S.~ Vagnozzi et al.,
\emph{Horizon-scale tests of gravity theories and fundamental physics from the Event Horizon Telescope image of Sagittarius A,}
Class. Quant. Grav. \textbf{40} no.16, (2023) 165007,
[arXiv:2205.07787 [gr-qc]].

\bibitem{DiGennaro:2022grw}
S.~Di Gennaro, H.~Xu and Y.~C.~Ong,
\emph{How barrow entropy modifies gravity: with comments on Tsallis entropy,}
Eur. Phys. J. C \textbf{82} no.11, (2022) 1066,
[arXiv:2207.09271 [gr-qc]].

\bibitem{DiGennaro:2022ykp}
S.~Di Gennaro and Y.~C.~Ong,
\emph{Sign Switching Dark Energy from a Running Barrow Entropy,}
Universe \textbf{8} (2022) 541,
[arXiv:2205.09311 [gr-qc]].

\bibitem{Nojiri:2022aof}
S.~Nojiri, S.~D.~Odintsov and V.~Faraoni,
\emph{From nonextensive statistics and black hole entropy to the holographic dark universe,}
Phys. Rev. D \textbf{105} no.4, (2022) 044042,
[arXiv:2201.02424 [gr-qc]].

\bibitem{Nojiri:2022dkr}
S.~Nojiri, S.~D.~Odintsov and T.~Paul,
\emph{Early and late universe holographic cosmology from a new generalized entropy,}
Phys. Lett. B \textbf{831} (2022) 137189,
[arXiv:2205.08876 [gr-qc]].

\bibitem{Nutma:2013zea}
T.~Nutma,
\emph{xTras : A field-theory inspired xAct  package for mathematica,}
Comput. Phys. Commun. \textbf{185} (2014) 1719--1738,
[arXiv:1308.3493 [cs.SC]].

\bibitem{Xiao:2021zly}
Y.~Xiao and Y.~Tian,
\emph{Logarithmic correction to black hole entropy from the nonlocality of quantum gravity,}
Phys. Rev. D \textbf{105} no.4, (2022) 044013,
[arXiv:2104.14902 [gr-qc]].

\bibitem{Abreu:2024tdv}
E.M.C.~Abreu,
\emph{Surface gravity analysis in Gauss-Bonnet and Barrow black holes,}
[arXiv:2403.02540 [gr-qc]].

\bibitem{Nojiri:2022sfd}
S.~Nojiri, S.~D.~Odintsov and V.~Faraoni,
\emph{Alternative entropies and consistent black hole thermodynamics,}
Int. J. Geom. Meth. Mod. Phys. \textbf{19}  no.13, (2022) 2250210,
[arXiv:2207.07905 [gr-qc]].



\end{thebibliography}
\end{document}